\documentclass{article}

\usepackage[T1]{fontenc} 
\usepackage[utf8]{inputenc} 
\usepackage{ismir,amsmath,cite,url}
\usepackage{graphicx}
\usepackage{color}

\DeclareMathOperator*{\argmax}{arg\,max}

\title{Melody transcription via generative pre-training}

\threeauthors
  {Chris Donahue} {Stanford University}
  {John Thickstun} {Stanford University}
  {Percy Liang} {Stanford University}

\def\authorname{C. Donahue, J. Thickstun, and P. Liang}

\usepackage[bookmarks=false,pdfauthor={\authorname},pdfsubject={\papersubject},hidelinks]{hyperref}

\usepackage{booktabs}
\usepackage{cleveref}
\usepackage{amsfonts}
\usepackage{amssymb}

\newcommand{\madmom}{\texttt{madmom}}
\newcommand{\mel}{Mel}
\newcommand{\mtthree}{MT3}
\newcommand{\jukebox}{Jukebox}
\newcommand{\hooktheory}{HookTheory}
\newcommand{\rwc}{RWC-MDB}
\newcommand{\sheetsage}{Sheet Sage}
\newcommand{\fone}{\texttt{F1}}
\newcommand{\fnot}{F0}
\newcommand{\Beatpooling}{Beat-wise resampling}
\newcommand{\beatpooling}{beat-wise resampling}

\begin{document}

\maketitle

\begin{abstract}
Despite the central role that melody plays in music perception, 
it remains an open challenge in MIR to reliably detect the notes of the melody present in an arbitrary music recording. 
A key challenge in \emph{melody transcription} is building methods which can handle broad audio containing any number of instrument ensembles and musical styles---existing strategies work well for \emph{some} melody instruments or styles but not all. 
To confront this challenge, we leverage representations from \jukebox{}~\cite{dhariwal2020jukebox}, 
a generative model of broad music audio, 
thereby improving performance on melody transcription 
by 
$20$\% 
relative to conventional spectrogram features. 
Another obstacle in melody transcription is a lack of training data---we derive a new dataset containing $50$ hours of melody transcriptions from crowdsourced annotations of broad music. 
The combination of generative pre-training and a new dataset for this task results in 
$77\%$ stronger performance on melody transcription relative to the strongest available baseline.\footnote{Examples: \url{https://chrisdonahue.com/sheetsage} \\
Code: \url{https://github.com/chrisdonahue/sheetsage}
\label{sound_examples}} 
By pairing our new melody transcription approach with solutions for beat detection, key estimation, and chord recognition, 
we build Sheet Sage, a system capable of transcribing human-readable lead sheets directly from music audio.
\end{abstract}

\section{Introduction}\label{sec:introduction}

In the Western music canon, 
\emph{melody} is a defining characteristic of musical composition, 
and can even constitute the very identity of a piece of music within the collective consciousness. 
Because of the significance of melody to our music perception, 
the ability to automatically transcribe the melody notes present in an arbitrary recording 
could enable numerous applications in 
interaction~\cite{ryynanen2008accompaniment}, 
education~\cite{droe2006music}, 
informatics~\cite{bainbridge1999towards}, 
retrieval~\cite{ghias1995query}, 
source separation~\cite{ewert2014score},
and generation~\cite{hawthorne2019enabling}.
Despite the potential benefits, 
reliable melody transcription remains an open challenge.

A closely-related problem that has received considerable attention from the MIR community is \emph{melody extraction}~\cite{goto1999real,goto2004real,salamon2014melody,rao2022melody}, where the goal is to estimate the time-varying, continuous \fnot{} trajectory of the melody in an audio mixture. 
In contrast, the goal of melody transcription is to output the \emph{notes} of the melody, where a note is defined by an onset time, a pitch, and an offset time. 
While \fnot{} trajectories are useful for several downstream tasks (e.g.,~query by humming) and more inclusive of music which does not use equal-tempered pitches, unlike notes, trajectories cannot be readily converted into formats like MIDI or scores which are more convenient for musicians.

\begin{figure}
    \centering
    \includegraphics[width=8.1cm]{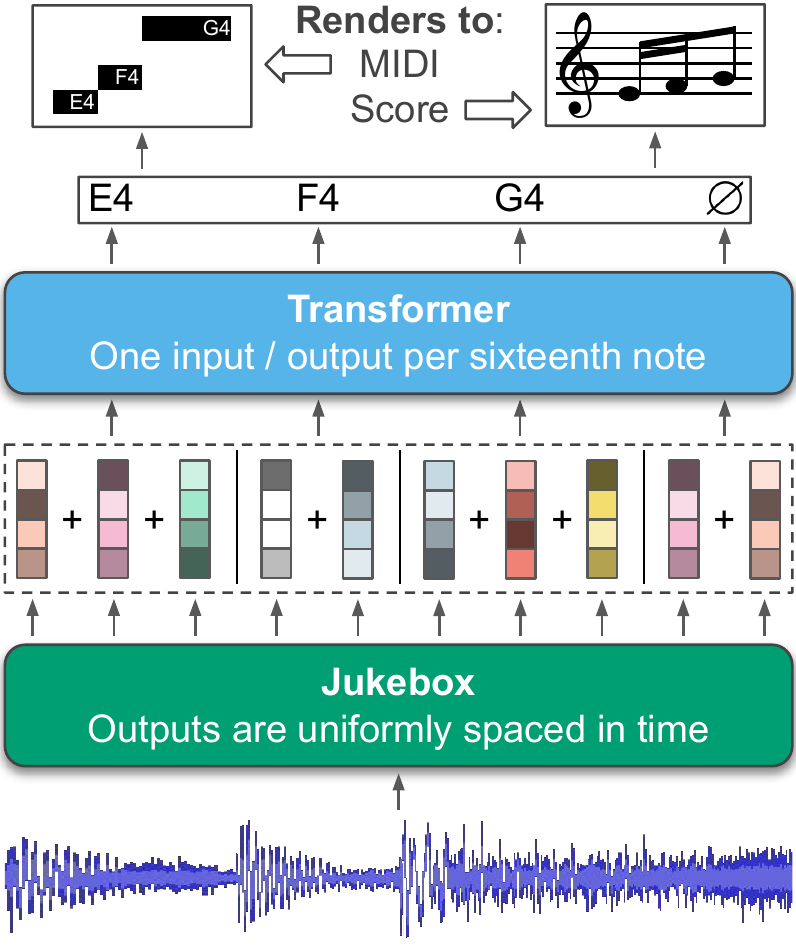}
    \caption{
Our melody transcription approach involves 
(1)~extracting audio representations from Jukebox~\cite{dhariwal2020jukebox}, a generative model of music, 
(2)~averaging these representations across time to their nearest sixteenth note (dashed outline---uses \madmom{}~\cite{bock2016joint,bock2016madmom} for beat detection),
and
(3)~training a Transformer~\cite{vaswani2017attention} to detect note onsets (or absence thereof) per sixteenth note. 
Outputs can be rendered to MIDI (by mapping beats back to time) or a score.
}
 \label{fig:fig1}
 \vspace{-5mm}
\end{figure}

The relative lack of progress on melody transcription is perhaps counterintuitive when compared to the considerable progress on seemingly more difficult tasks like piano transcription~\cite{sigtia2016end,hawthorne2017onsets}.
This circumstance stems from two primary factors.
First, unlike in piano transcription, melody transcription involves operating on \emph{broad}
audio mixtures from arbitrary instrument ensembles and musical styles. 
Second, there is a deficit of training data for melody transcription, which particularly impedes the deep learning approaches central to recent improvements on other transcription tasks. 
Moreover, collecting data for melody transcription is difficult compared to collecting data for tasks like piano transcription, where a Disklavier can be used to create aligned training data in real time. 

To overcome the challenge of transcribing broad audio, in this work we leverage representations from Jukebox~\cite{dhariwal2020jukebox}, a large-scale generative model of music audio pre-trained on $1$M songs. 
In~\cite{castellon2021calm}, Castellon~et~al.\ demonstrate that representations from Jukebox are useful for improving performance on a wide variety of MIR tasks. 
Here we show that, when used as input features to a Transformer model~\cite{vaswani2017attention}, representations from Jukebox yield 
$27$\% stronger performance on melody transcription (as measured by note-wise F1) relative to handcrafted spectrogram features conventionally used for transcription. 
To our knowledge, this is the first evidence that representations learned via generative modeling are useful for time-varying MIR tasks like transcription, as opposed to the song-level tasks (e.g.~tagging, genre detection) examined in~\cite{castellon2021calm}.

To address the data deficit for melody transcription, 
we release a new dataset containing $50$ hours of melody annotations for broad audio 
which we derive 
from \hooktheory.\footnote{\url{https://www.hooktheory.com/theorytab}} 
The user-specified alignments between the audio and melody annotations in \hooktheory{} are crude---we refine these alignments using beat detection. 
To overcome remaining alignment jitter, we resample features to be uniformly spaced in beats (rather than time) and pass these \emph{beat-wise resampled} features as input to melody transcription models. 
This procedure 
has a secondary benefit of enabling simple conversion from raw model outputs to human-readable scores (\Cref{fig:fig1}).

By training Transformer models on this new dataset using representations from Jukebox as input, we are able to improve overall performance on melody transcription by 
$70$\% relative 
to the strongest available baseline. 
A summary of our primary \textbf{contributions} follows:
\begin{itemize}
    \item We show that representations from generative models can improve melody transcription (\Cref{sec:experiments}).
    \item We collect, align, and release a new dataset with $50$ hours of melody and chord annotations (\Cref{sec:dataset}).
    \item We propose a method for training transcription models on data with imprecise alignment (\Cref{sec:beatpool}).
    \item As a bonus application of our melody transcription approach, we build a system which can transcribe music audio into lead sheets (\Cref{sec:sheetsage}).
\end{itemize}

\section{Related work}\label{sec:related}

Melody transcription is closely related to but distinct from the task of melody extraction, originally referred to as predominant fundamental frequency (\fnot) estimation~\cite{goto1999real,goto2004real}. 
Melody extraction has received significant interest from the MIR community over the last two decades (see~\cite{salamon2014melody,rao2022melody} for comprehensive reviews), 
and is the subject of an annual MIREX competition~\cite{downie2014ten}. 
Melody extraction may be a component of a melody transcription pipeline in combination with a strategy to segment \fnot{} into notes~\cite{salamon2015midi,nishikimi2016musical,nishikimi2017scale}---we directly compare to such a pipeline in \Cref{sec:exp2}.

Compared to melody extraction, melody transcription has received considerably less attention. 
Earlier efforts use sophisticated DSP-based pipelines~\cite{paiva2004auditory,paiva2005detection,ryynanen2008automatic,weil2009automatic}---unfortunately none of these methods provide code, though~\cite{ryynanen2008automatic} provides example transcriptions which we use to facilitate direct comparison. 
A more recent effort uses ground truth chord labels as extra information to improve melody transcription~\cite{laaksonen2014automatic}---in contrast, our method does not require extra information. 
Another line of work seeks to transcribe solo vocal performances into notes~\cite{molina2014sipth,mauch2015computer,nishikimi2020bayesian,nishikimi2021audio}. 
As singing voice often carries the melody in popular music, we directly compare to a baseline which firsts isolates the vocals (using Spleeter~\cite{hennequin2020spleeter}) and then transcribes them.

Polyphonic music transcription is another related task which involves transcribing \emph{all} of the notes present in a recording (not just the melody).
This task has its own MIREX contest (Multiple Fundamental \fnot{} Estimation) alongside a growing collection of supervised training data resources \cite{benetos2013automatic,thickstun2017learning,hawthorne2019enabling,manilow2019cutting}. 
The similarity of the polyphonic and melody transcription problems motivates us to experiment with representations learned by a polyphonic system---specifically, MT3 \cite{gardner2021mt3}---for melody transcription.

\section{Task definition}
\label{sec:task}

In this work, melody transcription refers to the task of converting a music recording into a \emph{monophonic} (non-overlapping) sequence of \emph{notes} which constitute its dominant melody.\footnote{Melody is difficult to precisely define---here we adopt an implicit definition based on a dataset of crowdsourced melody annotations.} 
More precisely, given a music waveform $\bm{a}$ of length $T$ seconds, our task is to 
uncover 
the sequence of $N$ notes~${\bm{y} = [\bm{y}_1,\dots,\bm{y}_N}]$ that represent the melody of $\bm{a}$.  For many MIR tasks, including transcription, it can be convenient to work with \emph{features} of audio ${\bm{X} = \texttt{Featurize}(\bm{a})}$, rather than waveforms. 
Hence, a melody transcription algorithm is a procedure that maps featurized audio to notes, i.e.~${\bm{y} = \texttt{Transcribe}(\bm{X})}$. 


Canonically, a musical note consists of an onset time, a musical pitch, and an offset time. 
However, in this work  
we disregard offsets and define a note to be a pair $\bm{y}_i = (t_i,n_i)$ consisting of an onset time~${t_i \in [0,T)}$ and discrete musical pitch~${n_i \in \mathbb{V} = \{\text{A0}, \ldots, \text{C8}\}}$.
We ignore offsets for 
two reasons. 
First, accurate offsets have been found to be considerably less important for human perception of transcription quality compared to accurate onsets~\cite{ycart2020investigating}. 
Second, in our dataset,
a heuristically-determined offset is identical to the user-annotated offset for $89\%$ of notes.\footnote{The specific heuristic that we use sets the offset of one note equal to the onset of the next, i.e., it assumes the melody is legato.}

Formally, a musical audio recording of length~$T$ seconds sampled at rate~$f_s$ is a vector~${\bm{a} \in \mathbb{R}^{Tf_s}}$. 
A featurization of audio ${\bm{X} \in \mathbb{R}^{Tf_k \times d}}$ is a matrix of $d$-dimensional features of audio, sampled uniformly at some rate ${f_k \ll f_s}$ (for example, $\bm{X}$ could be a spectrogram).
Intuitively, the function ${\texttt{Featurize} : \mathbb{R}^{Tf_s} \to \mathbb{R}^{Tf_k \times d}}$ defined by ${\bm{a} \mapsto \bm{X}}$ maps 
raw audio to a feature representation more conducive to learning. 
A melody of length $N$ is a sequence of notes 
$\bm{y} = [\bm{y}_1,\dots,\bm{y}_N] \in \mathbb{Y}^N$
consisting of onset-pitch pairs ${\bm{y}_i = (t_i,n_i)} \in \mathbb{Y} = \mathbb{R}^+ \times \mathbb{V}$ where ${t_i < t_j}$ if ${i < j}$. Given a featurization $\bm{X}$, the melody transcription task is to construct a transcription algorithm ${\texttt{Transcribe} : \mathbb{R}^{Tf_k \times d} \to \mathbb{Y}^N}$ such that ${\bm{X} \mapsto \bm{y}}$.

\subsection{Evaluation}
\label{sec:eval}

To evaluate a melody transcription method $\texttt{Transcribe}$, 
we adopt a standard metric commonly used for evaluation in polyphonic music transcription tasks, namely,  ``onset-only note-wise F-measure''~\cite{ycart2020investigating}. 
This metric scores an estimated transcript $\texttt{Transcribe}(\bm{X})$ by first matching its note onsets to those in the reference $\bm{y}$ with $50$ms of tolerance (default in~\cite{raffel2014eval}), and then computes a standard \fone{} score where an estimated note is treated as correct if it is the same pitch as its matched reference note. 
This ``note-wise'' metric represents a departure from the ``frame-based'' metrics typically used to evaluate melody extraction algorithms---Ycart~et~al.\ demonstrate in~\cite{ycart2020investigating} that this particular note-wise metric correlates more strongly with human perception of transcription quality than any other common metric, including frame-based ones.

We make a slight modification to this note-wise metric 
specific to the melody transcription setting: an estimate $\texttt{Transcribe}(\bm{X})$ may receive full credit if it is off by a fixed octave shift but otherwise identical to the reference. 
In downstream settings, melody transcriptions are likely to be used in an octave-invariant fashion, e.g.,~they may be shifted to read more comfortably in treble clef, or performed by singers with different vocal ranges. 
Hence, we modify the evaluation criteria by simply taking the highest score over octave shifted versions of the estimate:
\begin{equation*}
    \max_{\sigma \in \mathbb{Z}} \texttt{\fone}(\texttt{OctaveShift}(\texttt{Transcribe}(\bm{X}), \sigma), \mathbf{y}).
\end{equation*}
Henceforth, we refer to this octave-invariant metric as \fone. 

\vspace{-2mm}
\section{Dataset overview}
\label{sec:dataset}

A major obstacle to progress on melody transcription is the lack of a large volume of data for training. 
To the best of our knowledge, 
there are only two datasets available with annotations suitable for melody transcription: the RWC Music Database~\cite{goto2002rwc,goto2003rwc,goto2004development} (RWC-MDB), 
and a dataset labeled by Laaksonen~\cite{laaksonen2014automatic}. 
The former is larger but the annotations are inconsistent---Ryyn{\"a}nen~and~Klapuri note that only $8.7$ hours ($130$ songs) are usable for melody transcription~\cite{ryynanen2008automatic}, while the latter only contains $1.5$ hours. 

We derive a suitably large dataset for melody transcription using crowdsourced annotations from \hooktheory{}.\footnote{
\hooktheory{} annotations are published under a \href{https://creativecommons.org/licenses/by-nc-sa/3.0/}{CC BY-NC-SA 3.0} license, which our dataset inherits.
}
\hooktheory{} is a platform where users can easily create and share musical analyses of particular recordings hosted on YouTube, with Wikipedia-style editing. 
The dataset contains annotations for $22$k segments of $13$k unique recordings totaling $50$ hours of labeled audio. 
The audio content covers a wide range of genres---there is a skew towards pop and rock but many other genres are represented including EDM, jazz, and even classical. 
We create an artist-stratified $8$:$1$:$1$ split of the dataset for training, validation, and testing. 
The dataset also includes chord annotations which may facilitate chord recognition research.

While \hooktheory{} data has been used previously for MIR tasks like 
harmonization~\cite{chen2021surprisenet,yeh2021automatic}, 
chord recognition~\cite{jiang2019mirex}, and 
representation learning~\cite{jiang2020transformer}, 
making use of this platform for MIR is currently cumbersome. 
One obstacle is that the annotations are created via a ``functional'' interface, i.e.,~one which uses scale degrees and roman numerals relative to a key signature instead of absolute notes and chord names. 
In contrast, most MIR research favors absolute labels.
Hence, we convert annotations from this functional format to a simple (JSON-based) absolute format. 
One caveat is that the \hooktheory{} annotation interface uses a relative octave system, 
so there is no way to reliably map annotations to a ground truth octave.
Thus, melodies in our dataset also contain only relative octave information, consistent with the octave-invariant evaluation proposed in \Cref{sec:eval}.

\vspace{-3mm}
\section{Methods}

Similar to state-of-the-art methodology used for polyphonic transcription~\cite{hawthorne2021sequence}, 
our approach to melody transcription involves training Transformer models~\cite{vaswani2017attention} to predict notes from audio features. 
However, to address the unique challenges of melody transcription, our approach differs in two distinct ways. 
First, because melody transcription involves operating on broad audio, we leverage representations from pre-trained models as drop-in replacements for the handcrafted spectrogram features used as inputs to other transcription systems. 
Secondly, because alignments in our dataset are approximate, we propose a new strategy for training transcription models under such conditions.

\subsection{Pre-trained representations}
\label{sec:representations}

We explore representations from two different pre-trained models for use as input features to transcription models.
In~\cite{castellon2021calm}, Castellon~et~al.\ demonstrate that representations from \jukebox~\cite{dhariwal2020jukebox}---a generative model of music audio pre-trained on $1$M songs---constitute effective features for many MIR tasks, though notably they do not experiment on transcription. 
We adopt their approach to extract features from \jukebox{} (${f_k \approx 345}$~Hz,~${d = 4800}$), though we use a deeper layer~($53$) than their default~($36$) which improved transcription performance in our initial experiments. 

We also explore features from \mtthree~\cite{gardner2021mt3}, an encoder-decoder transcription model pre-trained on a multitude of different transcription tasks (though not melody transcription). 
For this model, we use the encoder's outputs as features (${f_k = 125}$~Hz,~${d = 512}$). 
The two models have different trade-offs with respect to our setting: \jukebox{} was pre-trained on audio similar to that found in our dataset but in a generative fashion, 
whereas \mtthree{} is pre-trained on transcription but for different audio domains.

\subsection{Refined Alignments}
\label{sec:align}

\begin{figure}
    \centering
    \includegraphics[width=8.1cm]{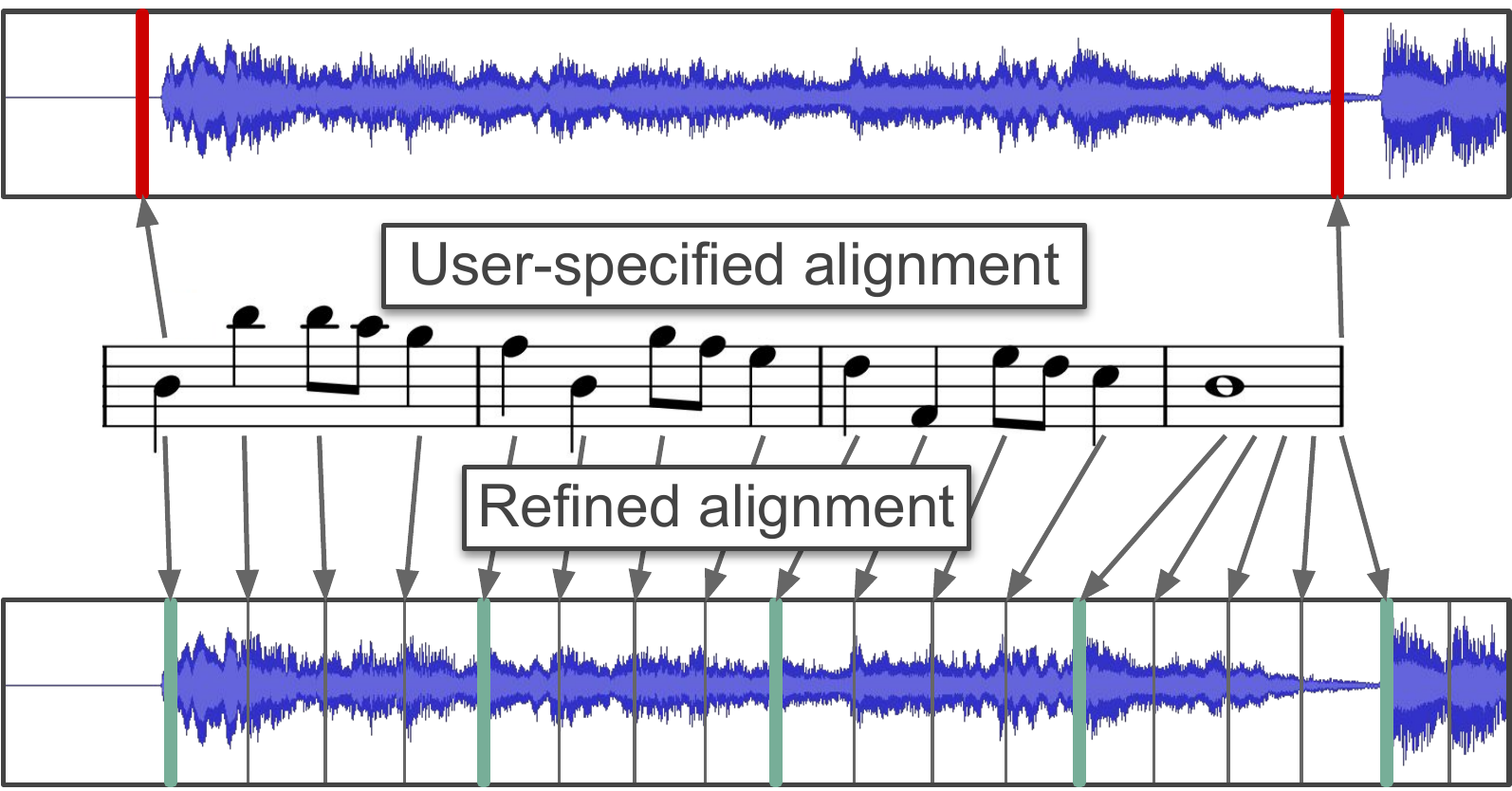}
    \caption{We refine the crude user-specified alignments from \hooktheory{} by using beat and downbeat tracking. The first segment beat is mapped to the detected downbeat nearest to the user-specified starting timestamp, and remaining beats are mapped to subsequent detected beats.}
 \label{fig:alignment}
 \vspace{-5mm}
\end{figure}

The alignments between audio and HookTheory annotations are crude---users provide only an approximate starting and ending timestamp of their annotated segment within the audio. 
Because transcription methodology generally depends on precise alignments, we make an effort to refine the user-specified ones. 
To this end, we make use of the beat and downbeat detection algorithm from \madmom{}~\cite{bock2016joint,bock2016madmom}. 
Specifically, our approach aligns the first beat of the segment to the detected downbeat which is nearest to the user-specified starting timestamp. 
Then, we align the remaining beats to the subsequent detected beats (see~\Cref{fig:alignment} for an example). 
This provides a beat-level alignment for the entire segment, which we linearly interpolate to fractional subdivisions of the beat. 
Formally, we construct an alignment function $\texttt{Align} : [0,B) \to [0,T)$ that assigns each of $B$ beats in the metrical structure to a time $t \in [0,T)$ in the audio.
In an informal listening test, this produced an improved alignment for $95$ of $100$ segments, 
where the primary failure mode in the remaining $5$ segments occurred when \madmom{} detected the wrong beat as the downbeat. 
We use these refined alignments for training and evaluation and release them alongside the dataset.

\subsection{\Beatpooling}
\label{sec:beatpool}

Here we outline our approach for training transcription models in the presence of imprecise alignments. 
Existing transcription methods were largely designed for domains where perfect alignments are readily available, e.g.,~piano transcription data captured by a Disklavier. 
Despite our best efforts, the refined \hooktheory{} alignments are still imprecise when compared to alignments in the datasets used to develop existing methods. 
Consequently, in initial experiments, we found that naively adopting existing methods (specifically, \cite{hawthorne2017onsets,hawthorne2021sequence}) resulted in poor performance on our dataset and task. 
Additionally, initial experiments on training models with an alignment-free approach~\cite{graves2006connectionist} also resulted in poor performance.

Accordingly, to sidestep small alignment deviations, we perform a \beatpooling{} of audio features $\bm{X}~\in~\mathbb{R}^{Tf_k \times d}$ to yield features that are uniformly spaced in subdivisions of the beat (using $\texttt{Align}$---see \Cref{sec:align}) rather than in time. 
For an audio recording with $B$ beats, we sample features $\tilde{\bm{X}} \in \mathbb{R}^{4B \times d}$ at sixteenth-note intervals. 
The value $\tilde{\bm{X}}_i$ is constructed by averaging all feature vectors in $\bm{X}$ that are nearest to the $i$'th sixteenth note into a single vector which acts as a proxy feature. 
For example, if a recording has a tempo of $120$~BPM, a sixteenth note represents $125$~ms of time, which would entail averaging across 
$43$ feature vectors from \jukebox{} (${f_k \approx 345}$ Hz). 
The intuition is that, while our alignments may not be precise enough to identify which of those $43$ frames contains an onset, we can be reasonably confident that it occurs \emph{somewhere} within them, and thus the relevant frame will be incorporated into the proxy. 
A similar approach was previously explored for song structure analysis in~\cite{mcfee2014analyzing}.


\subsection{Modeling}
\label{sec:modeling}

Together with the \beatpooling{} ${\tilde{\bm{X}} \in \mathbb{R}^{4B\times d}}$, we convert 
the sparse task labels ${\bm{y} \in (\mathbb{R}^+ \times \mathbb{V})^N}$ into 
a dense sequence  
${\tilde{\bm{y}} \in \{\{\varnothing\}\cup\mathbb{V}\}^{4B}}$,
which indicates whether or not an onset occurs at each sixteenth note.\footnote{This requires quantizing labels to the nearest sixteenth note. In practice, less than $1\%$ of notes in our dataset are affected by this quantization.} Formally, 
\[
\tilde{\bm{y}}_i =
\begin{cases}
n_j & \text{ if $\texttt{Align}(\frac{i}{4}) = t_j$ for some note $\bm{y}_j$}, \\
\varnothing & \text{ otherwise}.
\end{cases}
\]
We formulate melody transcription as an aligned sequence-to-sequence modeling problem and 
attempt to predict the sequence $\tilde{\bm{y}}$ given $\tilde{\bm{X}}$. Specifically, we train models of the form ${f_{\theta} : \mathbb{R}^{4B \times d} \to \mathbb{R}^{4B \times (|\mathbb{V}| + 1)}}$, which parameterize probability distributions  ${p_\theta(\tilde{\bm{y}}_i|\bm{\tilde{X}}) = \texttt{SoftMax}(f_{\theta}(\tilde{\bm{X}})_i)}$ over elements of the sequence $\tilde{\bm{y}}$.
One unique aspect of our dataset is that absolute octave information is absent (see \Cref{sec:dataset}). 
Hence, we construct an octave-tolerant cross-entropy loss by 
identifying the octave shift amount that minimizes the standard cross-entropy loss (denoted \texttt{CE}) when applied to the labels:
\begin{equation*}
\operatorname*{min}_{\sigma \in \mathbb{Z}} \sum_{i=0}^{4B-1} \texttt{CE}(p_\theta(\tilde{\bm{y}}_i|\bm{\tilde{X}}), \texttt{OctaveShift}(\tilde{\bm{y}}_i, \sigma)).
\end{equation*}

We require a thresholding scheme to convert the dense sequence of soft probability estimates $p_\theta(\tilde{\bm{y}}_i|\tilde{\bm{X}})$ into a sparse sequence of notes required by our task (see \Cref{sec:task}). Given a threshold $\tau\in\mathbb{R}$ (in practice, tuned on validation data), we define a sorted \emph{onset list}
\[
\mathcal{I} = \texttt{Sort}(\{i \in \{0,\dots,4B-1\} : p_\theta(\tilde{\bm{y}}_i = \varnothing|\tilde{\bm{X}}) < \tau\}).
\]
This should be interpreted as a list of $N$ metrical positions where an onset likely occurs. The timings of these onsets are given by the alignment, and we will predict the note-value with the highest probability. The sparse melody transcription is thus defined for $j=1,\dots,N$ by
\begin{align*}
&\texttt{Transcribe}(\tilde{\bm{X}})_j = (t_j,n_j),\text{ where }\\
&\quad t_j = \texttt{Align}\left(\frac{\mathcal{I}_j}{4}\right),\\
&\quad n_j = \argmax_{v\in\mathbb{V}} p_\theta(\tilde{\bm{y}}_{\mathcal{I}_j} = v|\tilde{\bm{X}}).
\end{align*}

\begin{table}[t]
    \centering
    \begin{tabular}{lcc}
\toprule
Features & $d$ & \fone{} \\
\midrule
\mel{} & $229$ & $0.514$ \\
\mtthree{} & $512$ & $0.550$ \\
\jukebox{} & $4800$ & $\bm{0.615}$ \\
\midrule
\mel{}, \mtthree{} & $741$ & $0.548$ \\
\mel{}, \jukebox{} & $5029$ & $0.617$ \\
\mtthree{}, \jukebox{} & $5312$ & $0.622$ \\
\mel{}, \mtthree{}, \jukebox{} & $5541$ & $\bm{0.623}$ \\
\bottomrule
    \end{tabular}
    \caption{\hooktheory{} test set performance for Transformers trained with different features (top) and combinations (bottom). Features are complementary---combining all three yields highest performance---but marginally so compared to \jukebox{} alone.}
    \label{tab:hooktheory_test}
    \vspace{-3mm}
\end{table}

\section{Experiments}
\label{sec:experiments}

Here we describe our experimental protocol for training melody transcription models on the \hooktheory{} dataset. 
The purpose of these experiments is two-fold. 
First, we compare representations from different pre-trained models to handcrafted spectrogram features to determine if pre-training is helpful for the task of melody transcription (\Cref{sec:exp1}). 
Second, we compare our trained models holistically to other melody transcription baselines (\Cref{sec:exp2}).

All transcription models are encoder-only Transformers with the default hyperparameters from~\cite{vaswani2017attention}, 
except that we reduce the number of layers from $6$ to $4$ to allow models to be trained on GPUs with $12$GB of memory. 
During training, we select random slices from the annotated segments of up to $96$ beats or $24$ seconds in length (whichever is shorter). 
We train using our proposed loss function from~\Cref{sec:modeling} and perform early stopping based on max \fone{} score across thresholds $\tau$ on the validation set, using the best validation $\tau$ for testing. 
All models converge within $15$k steps or about a day on a single K40 GPU. 

\subsection{Comparing input features}
\label{sec:exp1}

We compare representations from \jukebox~\cite{dhariwal2020jukebox} and \mtthree~\cite{gardner2021mt3} (see~\Cref{sec:representations}) to handcrafted spectrogram features, 
which are commonly used by existing transcription methods.
Specifically, we compare to log-amplitude Mel spectrograms using the formulation from~\cite{hawthorne2017onsets} (${f_k \approx 31}$,~${d = 229}$). 
Because features may contain complementary information, we also experiment with all combinations of these three features. 
Note that our \beatpooling{} strategy allows for trivial combination of these features (by concatenation) despite their differing rates. 
In~\Cref{tab:hooktheory_test}, we report \fone{} (as described in~\Cref{sec:eval}) on the \hooktheory{} test set for all input features.

\begin{table}[t]
    \centering
    \begin{tabular}{lcc}
\toprule
Approach & \fone{} (All) & \fone{} (Vocal)\\
\midrule
MT3 Zero-shot~\cite{gardner2021mt3} & $0.133$ & $0.085$ \\
Melodia~\cite{salamon2012melody} + Segmentation & $0.201$ & $0.268$ \\
Spleeter~\cite{hennequin2020spleeter} + Tony~\cite{mauch2015computer} & $0.341$ & $\bm{0.462}$ \\
DSP + HMM~\cite{ryynanen2008automatic} & $\bm{0.420}$ & $0.381$ \\
\midrule
\mel{} + Transformer & $0.631$ & $0.621$ \\
\mtthree{} + Transformer & $0.701$ & $0.659$ \\
\jukebox{} + Transformer & $\mathbf{0.744}$ & $\mathbf{0.786}$ \\
\bottomrule
    \end{tabular}
    \caption{Performance of different approaches on a subset of \rwc~\cite{goto2002rwc,goto2003rwc,goto2004development}. The bottom three approaches were trained on the \hooktheory{} dataset. For fair comparison to vocal transcription baselines, we also separately report performance on the vocal portions of this dataset.}
    \label{tab:rwc_ryy}
    \vspace{-4mm}
\end{table}

\begin{figure*}
    \centering
    \includegraphics[width=\linewidth]{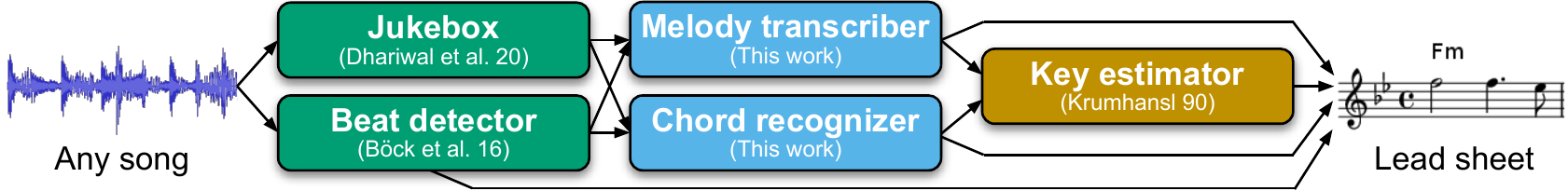}
    \caption{Inference procedure for Sheet Sage, our proposed system which transcribes any Western music audio into lead sheets (scores which depict melody as notes and harmony as chord names). The green, blue, and yellow boxes respectively take audio, features, and symbolic music data as input. Green boxes are modules that we built as part of this work---both are Transformers~\cite{vaswani2017attention} trained on their respective tasks using audio features from Jukebox~\cite{dhariwal2020jukebox} and data from \hooktheory~\cite{hooktheory}.}
    \label{fig:sheet_sage}
    \vspace{-3mm}
\end{figure*}

Overall, using representations from \jukebox{} as input features results in stronger melody transcription performance than using either representations from \mtthree{} or conventional handcrafted features. 
Representations from both \mtthree{} and \jukebox{} outperform conventional handcrafted features, 
implying that both pre-training strategies are helpful for melody transcription. 
Note that these two pre-training approaches are compared holistically---these models differ on several axes 
(number of parameters, 
pre-training data semantics, 
pre-training task), 
and thus it is impossible to disentangle the individual contributions of these different factors without retraining the models. 

Qualitatively speaking, there is a noticeable difference in performance across the three different input features which correlates with quantitative performance (see~\cref{sound_examples} for sound examples). 
Using representations from \jukebox{} tends to result in fewer wrong notes than the other features, and substantially reduces the number of egregiously wrong notes (e.g.,~notes outside of the key signature). 
Representations from \jukebox{} also appear to aid in the detection of more nuanced rhythmic patterns. 
Moreover, using handcrafted features will often result in several repeated onsets during a longer sustained melody note---in contrast, using representations from \jukebox{} appears to mitigate this failure mode.

Different features also appear to complement one another to a degree. 
The strongest performance overall is obtained by combining all three features, though the improvement over \jukebox{} alone is marginal. 
The practical downsides of combining all features outweigh the marginal benefits---running both pre-trained models effectively doubles the overall runtime, and the models have incompatible software dependencies. 
Hence, in the remainder of this paper we focus on models trained on individual features.

\vspace{-1mm}
\subsection{Comparison to melody transcription baselines}
\label{sec:exp2}

We compare overall performance of our proposed melody transcription approach to several baselines. 
We evaluate all methods on a small subset of $10$ songs from RWC-MDB~\cite{goto2002rwc,goto2003rwc,goto2004development}, 
another dataset which includes melody transcription labels. 
We chose this specific subset in an effort to compare to early DSP-based work on melody transcription---none of the early approaches~\cite{paiva2004auditory,paiva2005detection,ryynanen2008automatic,weil2009automatic} shared code, however~\cite{ryynanen2008automatic} shared melody transcriptions for this $10$-song subset.

In addition to~\cite{ryynanen2008automatic}, we also compare to a baseline which applies a note segmentation heuristic~\cite{salamon2015midi} to a melody extraction algorithm~\cite{salamon2012melody}. 
We additionally compare to \mtthree{} in a zero-shot fashion---this model was not trained on melody transcription but was trained on some tasks which incorporate vocal transcription. 
Finally, because the vocals often carry the melody in popular music, we compare to a baseline of running the Tony~\cite{mauch2015computer} monophonic transcription software on source-separated vocals isolated with Spleeter~\cite{hennequin2020spleeter}. 
Because this approach will only work for vocals, we also separately report performance on a subset of our evaluation set where the vocals represent the melody. 
Scores for all methods and baselines appear in~\Cref{tab:rwc_ryy}. 

Overall, our approach to training Transformers with features from \jukebox{} significantly outperforms the strongest baseline in both the vocals-only and unrestricted settings (${p < 0.01}$ using a two-sided t-test for paired samples). 
Qualitatively speaking, the stronger baselines produce transcriptions where a reasonable proportion of the notes are the correct pitches, but they have poor rhythmic consistency with respect to the ground truth. 
In contrast, our best model produces the correct pitches more often and with a higher degree of rhythmic consistency.

\vspace{-2mm}
\section{Sheet Sage}
\label{sec:sheetsage}

As a bonus demo, 
here we describe \sheetsage, a system we built to automatically convert music audio into lead sheets (see footnote on first page for examples), powered by our \jukebox-based melody transcription model. 
In Western music, a piece can often be characterized by its melody and harmony. 
When engraved as a lead sheet---a musical score containing the melody as notes on a staff and the harmony as chord names---melody and harmony can be readily interpreted by musicians, enabling recognizable performances of existing pieces. 
Hence, for some music, a lead sheet represents the essence of its underlying composition.
Existing services like Chordify~\cite{de2014chordify} can already detect a subset of the information needed to produce lead sheets (specifically, chords, beats, and keys) for broad music audio. 
However, despite past research efforts~\cite{ryynanen2008automatic,weil2009automatic}, no user-facing service yet exists which can convert broad music audio into lead sheets, presumably due to the poor performance of existing melody transcription systems.

To build \sheetsage, we also train a \jukebox-based chord recognition model on the \hooktheory{} data, using the same methodology that we propose for melody transcription (we simply replace the target vocabulary of onset pitches with one containing chord labels).
Passing audio through our \jukebox{}-based melody transcription and chord recognition models results in a score like format containing raw note names and chord labels per sixteenth note. 
Engraving this information as a lead sheet requires additional information: the key signature and the time signature. 
We estimate the former using the Krumhansl-Schmuckler algorithm~\cite{krumhansl1990cognitive,temperley1999key}, which takes the symbolic melody and chord information as input. 
For the latter, we use \madmom~\cite{bock2016madmom,bock2016joint}. 
Finally, we engrave a lead sheet using Lilypond~\cite{nienhuys2003lilypond}. 
See~\Cref{fig:sheet_sage} for a full schematic.

Subjectively speaking, \sheetsage{} often produces high-quality lead sheets, especially for the chorus and verse segments of pop music which have more prominent melodies. 
Performance is fairly robust across styles and instruments, even those which are less represented in the training data---one user reported particularly strong success on Bollywood music. 
However, the system occasionally struggles, especially with quieter vocals, layered harmonies, unusual time signatures, or poor intonation. 
\sheetsage{} is also limited to fixed time and key signatures due to limitations of its downbeat detection and key estimation modules.

\vspace{-2mm}
\section{Conclusion}

We present a new method and dataset which together improve melody transcription on broad music audio. 
Our method benefits from the rich representations learned by generative models pre-trained on broad audio. 
This suggests that further improvement in melody transcription may be possible without additional data, i.e.,~by scaling up or otherwise improving the pre-training procedure. 
By open sourcing our models and dataset, 
we hope to spark renewed interest for melody transcription in the MIR community, 
which may in turn reduce the gap between human perception and machine recognition of 
a fundamental aspect of music.


\vspace{-3mm}
\section{Ethical considerations}

Our definition of melody transcription incorporates equal temperament, a Western-centric tuning system. 
This could lead to disparate treatment of non equal-tempered music, e.g.,~if a streaming service were to use melody transcriptions for recommendation. 
We therefore advocate for the deployment of transcription only in contexts where users are self-selecting music to listen or play along to. 
Transcription may also be used to create training data for generation---as with any work on generation, there are risks of plagiarism and labor displacement. 
We recommend that any work on generation involve careful auditing and mitigation of plagiarism. 
Due to the incomplete nature of a melody, we argue that melody generation tools are more likely to be \emph{incorporated} into co-creation workflows (see~\cite{huang2020ai}) rather than used to displace musicians.

\section{Acknowledgements}

Thanks to 
Annie Hui-Hsin Hsieh, 
John Hewitt, 
Maggie Henderson, 
Megha Srivastava, 
Nelson Liu, 
Pang Wei Koh, 
Rodrigo Castellon, 
Sam Ainsworth, 
and
Zachary C. Lipton 
for helpful discussions, support, and advice. 
We thank all reviewers for their helpful comments.

\bibliography{main}

\end{document}